\documentclass[aps,prx,twocolumn,superscriptaddress,showpacs]{revtex4-1}

\usepackage{amsmath,amssymb,graphics,epsfig,epstopdf,color,verbatim,ulem,braket,tabularx}
\usepackage[colorlinks,linkcolor=blue,citecolor=blue,urlcolor=blue]{hyperref}

\begin{document}

\title{Caution on emergent continuous symmetry: a Monte Carlo investigation of transverse-field frustrated Ising model on the  triangle and the honeycomb lattices}
\author{Yan-Cheng Wang}
\affiliation{Beijing National Laboratory
for Condensed Matter Physics, and Institute of Physics, Chinese
Academy of Sciences, Beijing 100190, China}
\affiliation{School of Physical Sciences, University of Chinese Academy of Sciences, Beijing 100190, China}
\author{Yang Qi}
\affiliation{Perimeter Institute for Theoretical Physics, Waterloo, ON
  N2L 2Y5, Canada}
\affiliation{Department of physics, Massachusetts Institute of Technology, Cambridge, MA 02139, USA}
\author{Shu Chen}
\affiliation{Beijing National Laboratory for Condensed Matter Physics, and Institute of Physics, Chinese Academy of Sciences, Beijing 100190, China}
\affiliation{School of Physical Sciences, University of Chinese Academy of Sciences, Beijing 100190, China}
\author{Zi Yang Meng}
\affiliation{Beijing National Laboratory for Condensed Matter Physics, and Institute of Physics, Chinese Academy of Sciences, Beijing 100190, China}
\affiliation{School of Physical Sciences, University of Chinese Academy of Sciences, Beijing 100190, China}

\begin{abstract}
Continuous symmetries are believed to emerge at many
quantum critical points in frustrated magnets. In this work, we study two candidates of this paradigm: the transverse-field frustrated Ising model (TFFIM) on the triangle and the honeycomb lattices. The former is the prototypical example of this paradigm, and the latter has recently been proposed as another realization. Our large-scale Monte Carlo simulation confirms that the quantum phase transition (QPT) in the triangle lattice TFFIM indeed hosts an emergent O(2) symmetry, but that in the  honeycomb lattice TFFIM is a first-order QPT and does not have an emergent continuous symmetry. Furthermore, our analysis of the order parameter histogram reveals that such different behavior originates from the irrelevance and relevance of anisotropic terms near the QPT in the low-energy effective theory of the two models. The comparison between theoretical analysis and numerical simulation in this work paves the way for scrutinizing investigation of emergent continuous symmetry at classical and quantum phase transitions.
\end{abstract}


\date{\today} \maketitle

\section{INTRODUCTION}
Three score and seven years ago, Wannier~\cite{Wannier1950} and Houtappel~\cite{Houtappel1950} realized that two-dimensional (2D) antiferromagnetic Ising model on the triangle lattice does not order down to temperature $T=0$ -- in contrast to the naive expectation from the third law of thermodynamics -- thence initiated the study of frustrated magnetic system. By now, the scope of this field has been greatly expanded, where not only the aforementioned frustrated Ising model, but also models with continuous spin symmetry such as the antiferromagnetic Heisenberg~\cite{Yan2011} or XXZ magnets on 2D Kagome lattice~\cite{Isakov2011,Isakov2012,YCWang2017} and 3D pyrochlore lattice~\cite{Banerjee2008,Lv2015,CJHuang2017} are found to host exotic quantum disorder phases as their ground states, where novel phenomena such as topological orders~\cite{WenTO1990, Wen1991a, KivelsonPRB1987, WenZ2SL1991,YCWang2017}, emergent gauge fields~\cite{WenLeeSU2, arovas1988, HermeleU13D2004,CJHuang2017} and quantum phase transitions beyond the Landau-Ginzberg-Wilson (LGW) paradigm~\cite{senthil2004, ChubukovO4Star,Isakov2012,Lv2015} prevail.

Among the interesting phenomena associated with frustrated magnetic systems, the emergent continuous symmetry at the quantum critical point (QCP) in the transverse-field frustrated Ising model (TFFIM) holds a special position. In the by now canonical papers~\cite{Moessner2001,Isakov2003}, for the 2D TFFIM on the triangle lattice, as a function of magnetic field, the QCP between the magnetically ordered clock phase and the fully polarized phase are shown by both LGW renormalization group analysis and unbiased Monte Carlo (MC) simulations to be of the $(2+1)$D $O(2)$ universality class, despite the original Hamiltonian only contains discrete Ising symmetry. In a broader sense, emergent $O(n)$ symmetries have also been observed in the classical (finite temperature) transition in 3D $q$-state Potts model~\cite{Ding2015}, and in 2D, a QCP with an emergent $U(1)$ symmetry is the prominent feature of the famous deconfined quantum-critical point~\cite{Senthil2004a,Senthil2004b}, which separates antiferromagnetic N\'eel state and valence-bond-solid~\cite{Sandvik2007,Kaul2013,YQQin2017}.

The success of the theoretical prediction and numerical verification of the emergent $(2+1)$D $O(2)$ symmetry in the triangle lattice TFFIM~\cite{Moessner2001,Isakov2003} has bestowed confidence on people to find similar nontrivial QCPs in other models. However, one needs to be more cautious in generalizing the analysis to other systems. It is recently proposed that the TFFIM on the honeycomb lattice also hosts an emergent $(2+1)$D $O(3)$ continuous QCP~\cite{Roychowdhury2015}. However, this scenario can be destroyed by the cubic anisotropic perturbation, which may be a relevant perturbation at the (2+1)D $O(3)$ Wilson-Fisher fixed point~\cite{Calabrese2003} that renders this QPT first-order, as pointed out by the authors of Ref.~\cite{XuVBS2011}, who studied a similar possible QCP with an emergent $O(3)$ symmetry in a different model.

Here, by means of large-scale Monte Carlo simulations, we show that the two models -- the TFFIMs on the triangle and  the honeycomb lattices -- are in fact very different, that while the former indeed manifests an emergent QCP with $(2+1)$ $O(2)$ symmetry, the latter, unfortunately, hosts a first order quantum phase transition. The proposition of the emergent $(2+1)$D $O(3)$ symmetry in the TTFIM on the honeycomb lattice perishes and by exploiting the numerical simulation and data analysis to a higher level, we find out that the key difference between the previous theoretical analysis~\cite{Roychowdhury2015} and our numerical result is indeed the large and negative anisotropic terms in the effective LGW Hamiltonian, which are responsible for both the lack of an emergent continuous symmetry and the QPT being first-order. This is consistent with Ref.~\cite{XuVBS2011}, which points out that a negative cubic anisotropic term is relevant and will make the QPT first-order in their model.

The rest of the paper is organized as follows. In
Sec.~\ref{sec:model_method} the TFFIMs on the triangle and the honeycomb lattices (Sec.~\ref{sec:triangle_honeycomb}) and the Monte Carlo simulation techniques are introduced, with detailed accounts of the implementation of the efficient space-time cluster update scheme (Sec.~\ref{sec:QMC}) as well as the illustrative order parameter histogram method we developed here (Sec.~\ref{sec:orderparameterhisto}). In Sec.~\ref{sec:results}, the numerical results of the TFFIM on the triangle lattice (\ref{sec:results_triangle}) is first demonstrated, followed by those of the honeycomb lattice (\ref{sec:results_honeycomb}). In the case of the triangle lattice, the emergent $(2+1)$D $O(2)$ symmetry at the continuous QCP can be clearly seen from order parameter histogram and the Binder cumulant of magnetic moments. As for the honeycomb lattice, the order parameter histogram and the Binder cumulant analysis confirm the transition is of first order. In Sec.~\ref{sec:anisotropy}, we furthermore discover that the difference in the nature of the QPTs between the triangle and the honeycomb lattice models lies in the fact that the anisotropic term in the effective Lagrangian density is irrelevant/relevant in the former/latter. Hence, for the honeycomb lattice TFFIM, the presence of the anisotropic terms in the effective LGW Hamiltonian indicates that the previous field theoretical analysis~\cite{Roychowdhury2015} does not apply to this particular model. Section.~\ref{sec:summary} summarizes our findings.

\section{Models and numerical method}
\label{sec:model_method}

\subsection{Models}
\label{sec:triangle_honeycomb}
In this paper we study the TFFIM on the 2D triangle and honeycomb lattices~\cite{Blankschtein1984,Moessner2001,Isakov2003,Powalski2013,Roychowdhury2015}.

\begin{figure}[htp!]
\centering
\includegraphics[width=\columnwidth]{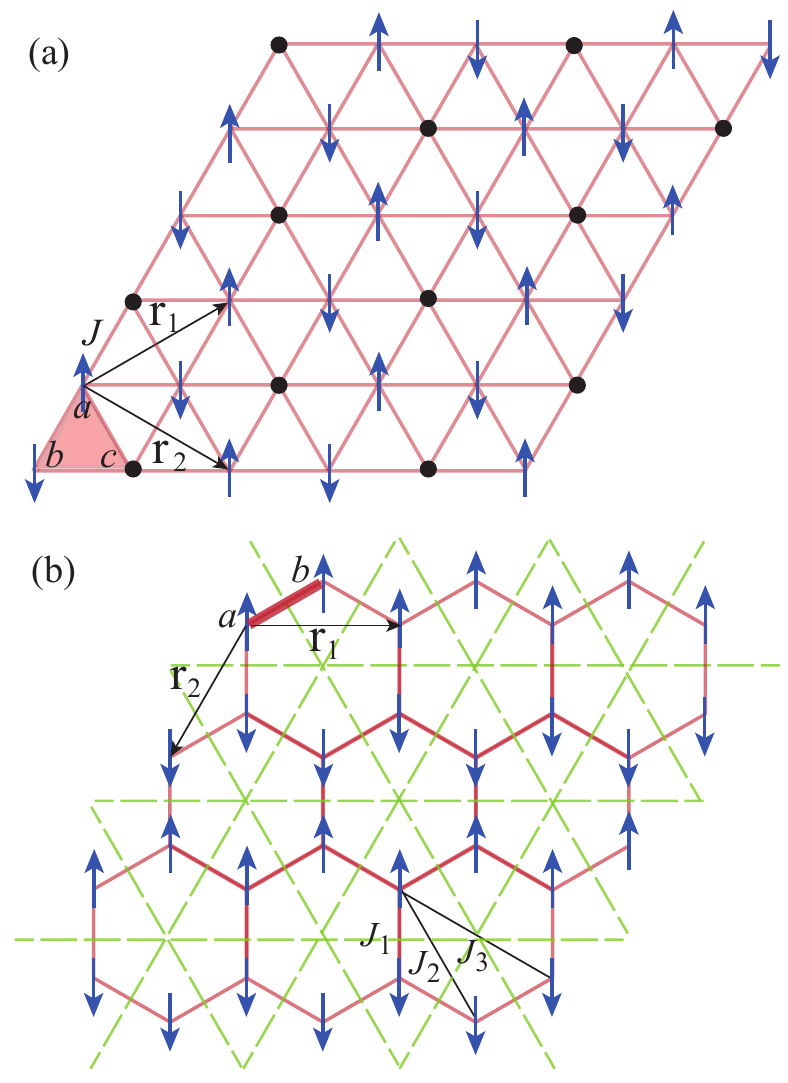}
\caption{(color online) (a). Triangular lattice. $a$, $b$ and $c$ are three sublattices. $J$ is the nearest neighbor antiferromagnetic coupling. $\mathbf{r}_1$ and $\mathbf{r}_2$ are the primitive vectors of the magnetically ordered clock phase. (b). Honeycomb lattice. $a$ and $b$ are the sublattices and $\mathbf{r}_1$ and $\mathbf{r}_2$ are the primitive vectors of the honeycomb lattice. $J_1$, $J_2$ and $J_3$ are the nearest, next-nearest and third-nearest neighbor antiferromagnetic couplings, respectively. The green dashed lines respect the $C_3$ rotational axes, along which the magnetically ordered phase are degenerate.}
\label{fig:lattices}
\end{figure}

The Hamiltonian for the TFFIM on the triangle lattice is given by
\begin{equation}
H = J\sum_{\langle i,j \rangle} \sigma_i^{z}\sigma_j^{z} - h\sum_{i}\sigma_i^{x},
\label{eq:triangle}
\end{equation}
where $J$ is the nearest-neighbor antiferromagnetic Ising coupling and $h$ is the transverse field. The three sublattice ($a$, $b$ and $c$) structure of the triangle lattice is given in Fig.~\ref{fig:lattices} (a), the spin orientation in Fig.~\ref{fig:lattices} (a) stands for the magnetically ordered clock phase~\cite{Blankschtein1984,Moessner2001,Isakov2003} when $h<h_c$, where $h_c$ is the QCP above which the system is fully polarized to $\sigma^{x}$ direction.

The Hamiltonian for the TFFIM on the honeycomb lattice is given as
\begin{equation}
{H}=J_1\sum_{\langle i,j\rangle}\sigma_i^z\sigma_j^z+J_2\sum_{\langle\langle i,j \rangle\rangle}\sigma_i^z\sigma_j^z
+J_3\sum_{\langle\langle\langle i,j \rangle\rangle\rangle}\sigma_i^z\sigma_j^z-h\sum_i \sigma_i^x
\label{eq:honeycomb}
\end{equation}
where $J_1$, $J_2$ and $J_3$ are the nearest, next-nearest and third-nearest neighbor antiferromagnetic couplings. The lattice structures and antiferromagnetic couplings for the honeycomb lattices are given in Fig.~\ref{fig:lattices} (b), the spin orientation in Fig.~\ref{fig:lattices} (b) stands for one of the six-fold degenerate magnetically ordered phase at $J_1=J_2=J_3$ and small $h$~\cite{Roychowdhury2015}. Throughout the paper, we set $J=J_1=1$ as the energy unit.

\subsection{Monte Carlo simulation}
\label{sec:QMC}
In a path-integral formalism, the 2D quantum ($T=0$) TFFIM can be mapped to a (2+1)D classical Ising model, where the Ising couplings in the time dimension are ferromagnetic while in the spatial dimensions are antiferromagnetic~\cite{Moessner2001,Isakov2003,BatrouniScalettar2011}. Taken Eq.~\ref{eq:triangle} as an example, the partition function can be expressed as

\begin{equation}
Z=\text{Tr} e^{-\beta H} = \text{Tr} \exp \{-\beta (J\sum_{\langle i,j \rangle} \sigma_{i}^{z}\sigma_{j}^{z}+h\sum_{i}\sigma_{i}^{x})\},
\end{equation}
where $\beta=\frac{1}{T}$. Within the Trotter-Suzuki formalism~\cite{BatrouniScalettar2011}, one can discretize the imaginary time axis into small pieces with footstep $\Delta\tau=\frac{\beta}{M}$ and $M\to\infty$, and the partition function is expressed as that of a $(2+1)$D classical system,
\begin{align}
Z &= \sum_{\{\sigma_{i}\}} \langle \{\sigma_{i}\} | \exp(-\Delta\tau H)^{M} | \{\sigma_{i}\}\rangle \nonumber\\
{} &= \prod_{l=1}^{M}\sum_{\{\sigma_{i,l}\}}\langle \{\sigma_{i,l}\} | \exp(-\Delta\tau H) |\{\sigma_{i,l+1}\}\rangle \nonumber\\
{} &= \prod_{l=1}^{M}\sum_{\{\sigma_{i,l}\}}\exp(-\Delta\tau J\sum_{\langle i,j \rangle}\sigma_{i}^{z}\sigma_{j}^{z}) \nonumber \\
 &\times \{\delta^{(0)}_{\{\sigma_{i,l},\sigma_{i,l+1}\}}+h\Delta\tau\delta^{(1)}_{\{\sigma_{i,l},\sigma_{i,l+1}\}}+O([\Delta\tau]^2Jh)\}
\label{eq:partition}
\end{align}
where the notation $\delta^{(k)}$ stands for unity if the two sets of spins consecutive in time differ by $k$ entries, and is equal to zero otherwise.

Eq.~\eqref{eq:partition} can be viewed as the partition function of a $(2+1)$D classical Ising system with (reduced) Hamiltonian
\begin{equation}
H = K\sum_{\langle i,j \rangle, l}\sigma_{i,l}^{z}\sigma_{j,l}^{z} - K^{\tau}\sum_{i,l}\sigma_{i,l}^{z}\sigma_{j,l+1}^{z},
\label{eq:d+1}
\end{equation}
where $K=J\Delta\tau$ and the effective Ising coupling in the time dimension is $K^{\tau}=-\frac{1}{2}\ln\tanh(\Delta\tau h)$. Such mapping becomes exact in the limit $\Delta\tau\to 0$ and $K^{\tau}\to\infty$. Technically speaking, such a limit will generate strong anisotropy in the coupling ratio $K^{\tau}/K$ and render the simulation very inefficient. Hence, to solve this problem, we design the combined MC update algorithm below.

To simulate the Hamiltonian in Eq.~\ref{eq:d+1}, we study the corresponding 3D classical Ising model using a Monte Carlo simulation. Although Metropolis local update scheme can be readily applied, in order to have an effective simulations, we use a combined algorithm which inlcudes local Metropolis update scheme, Wolff~\cite{Wolff1989} cluster update scheme, and the geometric cluster~\cite{Blote1995, Heringa1998} update scheme. In the cluster update schemes, we build cluster of sites in the $2+1$ space-time configuration space. The reason of employing such combined update scheme is that here to capture the QCP properly, we not only need to overcome the geometric frustration in spatial dimensions, but also need to beat the highly anisotropic coupling ratio $K^{\tau}/K \to \infty$ as $\Delta\tau\to0$. Moreover, the typical critical slowing down of Monte Carlo dynamics close to the QCP is also prominent and gives rise to many local minimals of the configuration space. Hence, only our combined space-time cluster scheme can overcome such three-fold difficulties while address the QCP in frustrated transverse field Ising models.

Each Monte Carlo step consists of three update steps: we first go through the $(2+1)$D space-time configuration 5 times with local Metropolis updates, then we try to construct the Wolff cluster over the lattice 5 times, note that the Wolff cluster has a tree structure which means in the case of the triangle lattice, from each lattice site, one tree has 8 branches (6 of them are in spatial dimension and the other 2 are in time dimension); whereas in the honeycomb lattice case, due to the frustrated $J_1$, $J_2$ and $J_3$ interactions, one site has 12 spatial neighbors so one tree can have 14 branches (12 of them are in spatial dimension and the rest 2 are in time dimension). ALL the spins associated with the space-time cluster are flipped. After the 5 Wolff cluster updates we also perform 5  geometric cluster updates. The three consecutive updates make sure that our spin configurations are sampled according to their Boltzmann weight, i.e., there is no ergodicity problem although our systems are highly frustrated, anisotropic and close to QCP.

The MC simulations are performed on lattice size of
$L=6,9,12,15,18$ for the triangle lattice and $L=6,8,10,12,16,20$ for the honeycomb lattice. We have tested that the convergence of the $\Delta\tau$, and find $\Delta\tau=0.02$ is sufficient for the accuracy requirement. And to obtain the ground state ($T=0$) properties in the thermodynamic limit, we scale $\beta=M \Delta\tau=2L$. For each simulation, we take about $5\times 10^5$ MC steps for equilibration and $5$ million MC steps for measurements.

\subsection{Order parameter histogram}
\label{sec:orderparameterhisto}
As shown in Ref.~\onlinecite{Blankschtein1984,Moessner2001,Isakov2003}, for the triangle lattice TFFIM, one can construct a complex $XY$ order parameter $\psi_{\pm}=m\exp(\pm i\theta)$, with two-fold degenerated ordered wave vector at the corner of the hexagonal Brillouin zone (BZ) $\mathbf{K}=(\frac{4\pi}{3},0)$ and $\mathbf{K'}=(-\frac{4\pi}{3},0)$. The corresponding LGW effective Hamiltonian is
\begin{equation}
H^{T}_{LGW} = \sum_{\mathbf{q}}(r+|\mathbf{q}|^2)m^2 + u_{4}m^4+u_6 m^{6} + \nu_{6}m^{6}\cos(6\theta).
\label{eq:effectiveHamiltonianT}
\end{equation}
The complex $XY$ order parameter deduced from the above LGW effective Hamiltonian can be measured and constructed from the MC simulation in the following way
\begin{equation}
me^{i\theta} \equiv (m_1+m_2 e^{i(4\pi/3)}+m_3 e^{i(-4\pi/3)})/\sqrt{3}
\label{eq:op-t}
\end{equation}
where $m_i$ $i=1,2,3$ are the sublattice magnetizations of the triangle lattice, as shown in Fig.~\ref{fig:lattices} (a).

\begin{figure}[h!]
\centering
\includegraphics[width=\columnwidth]{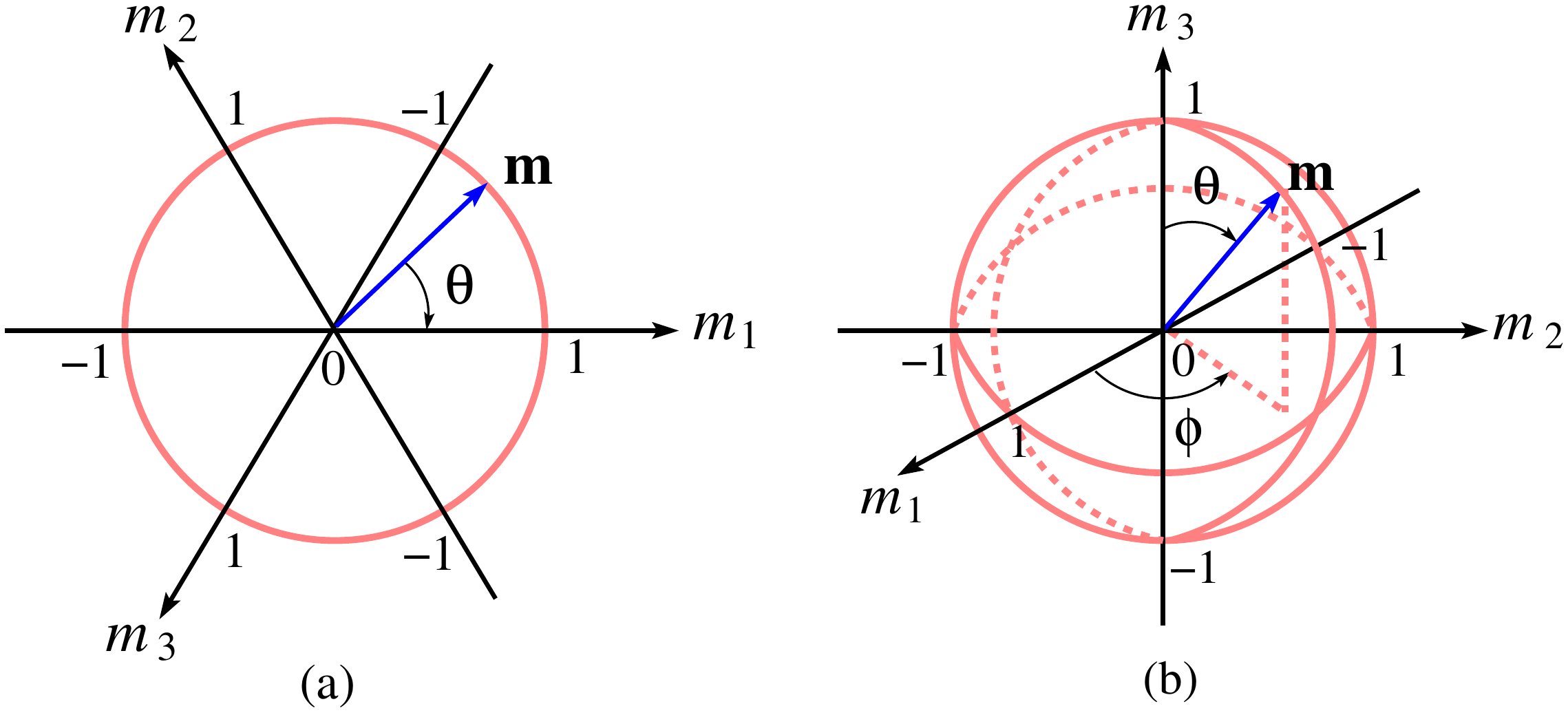}
\caption{(color online) (a). Chart for the order parameter of the triangle lattice TFFIM: $(m_1, m_2, m_3)$ are the three sublattice magnetization, separated by an angle of $\frac{2\pi}{3}$. (b). Chart for the order parameter of the honeycomb lattice TFFIM: $\vec{m}=(m_1, m_2, m_3)$ is a $O(3)$ vector of the three sublattice magnetization. It is presented in a spherical coordinate.}
\label{fig:op-th}
\end{figure}

To effectively illustrate the nature of the QCP in the triangle lattice TFFIM, i.e., whether there is emergent continuous $O(2)$ symmetry at the QCP, we designed the following order parameter histogram measurement: as shown in Fig.~\ref{fig:op-th}(a), the three axes of sublattice magnetization $m_1$, $m_2$ and $m_3$ are arranged into one chart, separated by an angle of $\frac{2\pi}{3}$. For each Monte Carlo configuration, a corresponding point with coordinate $(m_1, m_2, m_3)$ will be denoted in the chart. Over the Monte Carlo sampling process, a histogram of the distribution $(m_1, m_2, m_3)$ will be obtained. And since the Monte Carlo sampling process is performed according to the configuration weight in the partition function in Eq.~\ref{eq:partition}, such order parameter histogram can directly provide us the configuration distribution of the effective low-energy Hamiltonian in Eq.~\ref{eq:effectiveHamiltonianT}. In other word, from the order parameter histogram, we can directly observe which term would play the dominate role in the effective Hamiltonian in the vicinity of the QCP in the LGW Hamiltonian Eq.~\ref{eq:effectiveHamiltonianT}. As will be become clear in Sec.~\ref{sec:results}, this order parameter histogram turns out to be very powerful in revealing the nature of the QCP.

For the honeycomb lattice TFFIM, as discussed in Ref.~\onlinecite{Roychowdhury2015}, our choice of $J_{1}=J_2=J_3$ at $h<h_c$ gives rise an magnetically order ground state. The ordered wavevectors in the reciprocal space are located at the three inequivalent (with respect to reciprocal lattice vectors) $\mathbf{M}$ points of the hexagonal BZ: $\mathbf{M_{1}}=(\frac{\pi}{\sqrt{3}},-\frac{\pi}{3})$, $\mathbf{M_2}=(\frac{\pi}{\sqrt{3}},\frac{\pi}{3})$ and $\mathbf{M_3}=(0,\frac{2\pi}{3})$, which are related by rotational symmetry. As shown in Fig.~\ref{fig:lattices} (b), the ground state magnetic patterns in the ordered phase breaks the hexagon-centered six-fold rotational symmetry, and therefore are six-fold degenerate, similar to the triangular lattice TFFIM. (The ground states also break site-centered three-fold rotational symmetry, and the $\mathbb Z_2$ Ising symmetry.)


The LGW effective Hamiltonian of the transverse field honeycomb Ising model is given in Ref.~\cite{Roychowdhury2015}, it reads as
\begin{eqnarray}
H^{H}_{LGW} &=&\sum_{\mathbf{q}}(r+|\mathbf{q}|^2)m^2 + u_{4}m^4 + u_{6}m^6\nonumber\\
      &&+\nu_4(m_1^4+m_2^4+m_3^4)+\nu_6(m_1m_2m_3)^2,
\label{eq:h_lgw2}
\end{eqnarray}
where
\begin{equation}
m=|\mathbf{m}|=\sqrt{m_1^2+m_2^2+m_3^2},
\label{eq:op-h}
\end{equation}
is the length of a three-component vector. As shown in Fig.~\ref{fig:op-th}(b) , its components can be written in spherical coordinate as $m_1=m\sin\theta\cos\phi$, $m_2=m\sin\theta\sin\phi$,$m_3=m\cos\theta$. Different from the triangle lattice case, here $m_i$, $i=1,2,3$ stands for the magnetization of the patterns according to the $C_3$ rotational symmetry, as denoted by the green dashed lines in Fig.~\ref{fig:lattices}$(b)$. The order parameter histogram of the honeycomb lattice, can be performed as that of the triangle lattice aforementioned, with parameters in a $3$D unit sphere instead of the $2$D unit circle.

We would like to point out, that, the difference in the level of degeneracy for the magnetically ordered phase in TFFIM between the triangle lattice (at $\mathbf{K}$ and $\mathbf{K'}$ points) and the honeycomb lattice (at $\mathbf{M_1}$, $\mathbf{M_2}$ and $\mathbf{M_3}$ points), led to the proposal that the emergent continuous symmetry in the former is $O(2)$~\cite{Moessner2001,Isakov2003} and in the latter is $O(3)$~\cite{Roychowdhury2015}. In the next Section (Sec.~\ref{sec:results}), we will delineate the MC simulation results which confirm the emergent $O(2)$ symmetry at the QCP in the triangle lattice TFFIM, but disprove the emergence of the $O(3)$ symmetry at the QPT in the honeycomb lattice TFFIM. Furthermore, we will explain the reason behind such difference, in that, the anisotropic terms of the LGW effective Hamiltonian are irrelevant and vanishing in the former, but relevant and remain finite in the later.

\subsection{Binder cumulant}
In the study of magnetic phase transitions, the Binder cumulant is also a widely used observable. The normalized Binder cumulant \cite{Binder1981} for the triangle lattice is
\begin{equation}
U = 2\left(1-\frac{\langle m^4 \rangle}{2\langle m^2 \rangle^2}\right),
\label{eq:binder-t}
\end{equation}
and for the honeycomb case is,
\begin{equation}
U = \frac{5}{2}\left(1-\frac{3\langle m^4 \rangle}{5\langle m^2 \rangle^2}\right),
\label{eq:binder-h}
\end{equation}
where $m$ is the amplitude of the complex order parameters defined in Eq.~\ref{eq:op-t} and ~\ref{eq:op-h}. The Binder cumulant has a scaling dimension of zero. It thus has the advantage of not requiring fitting unknown leading exponents at the critical point and give unbiased information on position and nature of the QCP. The normalization factors are chosen in the way that when $L\to \infty$, the Binder cumulant  has the following behavior:
$U(L) \to 0$ at disorder phase, $U(L) \to 1$ at order phase and at the critical point $h_c$, $U(L)$ is becoming a step function.

Binder cumulant can also be used to identify the order of the phase transitions. As for a continuous phase transition, the Binder cumulant typically grows monotonically and stay bounded within $[0, 1]$, and it approaches a step function at $h_{c}$ in the thermodynamic limit~\cite{Binder1981}. But for a first order phase transition, it instead shows a nonmonotonic behavior with the control parameter for large systems~\cite{Vollmayr1993} -- developing a negative peak which approaches $h_c$ and grows narrower and diverges as $L^{2}$ when $L\to\infty$ in 2D system~\cite{Jin2012}. In the next section (Sec.~\ref{sec:results}), we indeed observe such difference in the Binder cumulant of the phase transition in the triangle and the honeycomb Ising TFFIMs.

\section{NUMERICAL RESULTS AND DISCUSSIONS}
\label{sec:results}

\subsection{Triangle lattice}
\label{sec:results_triangle}
\begin{figure}[htp!]
\centering
\includegraphics[width=\columnwidth]{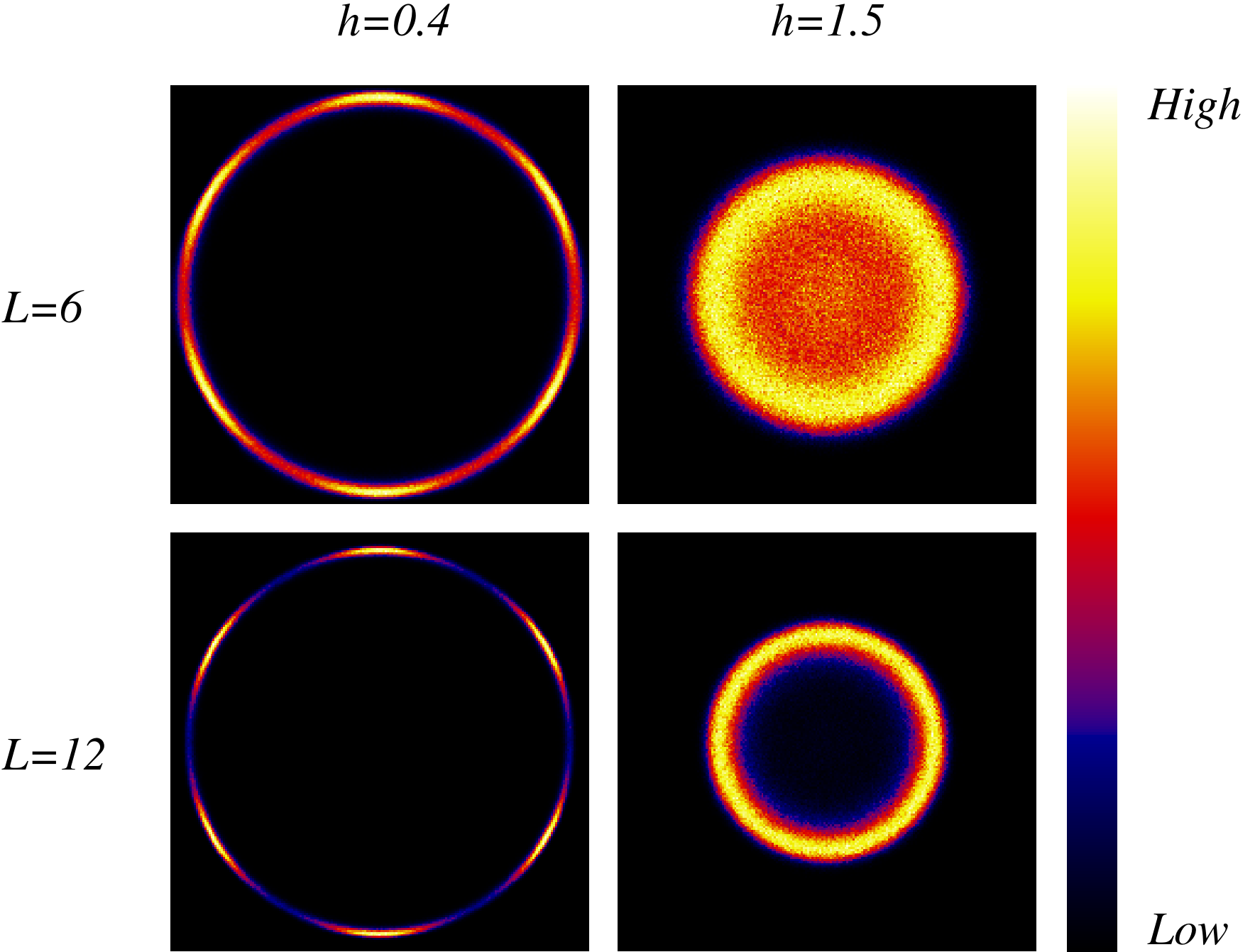}
\caption{(color online) Upper row: histograms of the order parameter for the triangle lattice TFFIM
at $h=0.4$ and $h=1.5$, with $L=6$. Lower row: histogram of the order parameter for the triangle lattice TFFIM with same set of $h$ for $L=12$. According to the crossing of the Binder cumulant in Fig.~\ref{fig:bindr-t}, the quantum critical point is at $h_c=1.64(1)$. Left panels show clearly 6-fold rotational symmetry and the right panels show emergent $O(2)$ symmetry close to the QCP.}
\label{fig:histogram_triangle}
\end{figure}

The emergent $O(2)$ symmetry at the critical point $h_c$ for the triangle lattice TFFIM has been investigated in several previous works~\cite{Blankschtein1984,Moessner2001,Isakov2003,Powalski2013}. The effective LGW Hamiltonian is given as Eq.~\ref{eq:effectiveHamiltonianT}. Fig.~\ref{fig:histogram_triangle} shows the order parameter histograms for the transverse field triangle lattice Ising model as defined in Eq.~\ref{eq:op-t}. The upper row is for smaller system with $L=6$ and the lower row is for larger system with $L=12$. At small transverse field ($h=0.4$), where the system is still in the ordered clock phase, the order parameter histogram is clearly inhomogeneous along the unit circle, as there are six bright arcs around $\theta=\frac{(2n+1)\pi}{6}$ with $n=0,1,2,3,4,5$. These six bright arcs correspond to the fact that deep in the clock phase, the anisotropic term in the LGW Hamiltonian, $\nu_6$, is finite and it dominates over the other terms, so the system is in a discrete symmetry breaking phase. However, as $h$ increases, the fluctuation of the angle $\theta$ becomes larger, and the order parameter histogram turns out to be a homogeneous ring (see the $h=1.5$ results in the right panels), i.e., the configuration weight of the order parameter starts to show a continuous $U(1)$ ($O(2)$) symmetry along the unit circle.

\begin{figure}[htp!]
\includegraphics[width=\columnwidth]{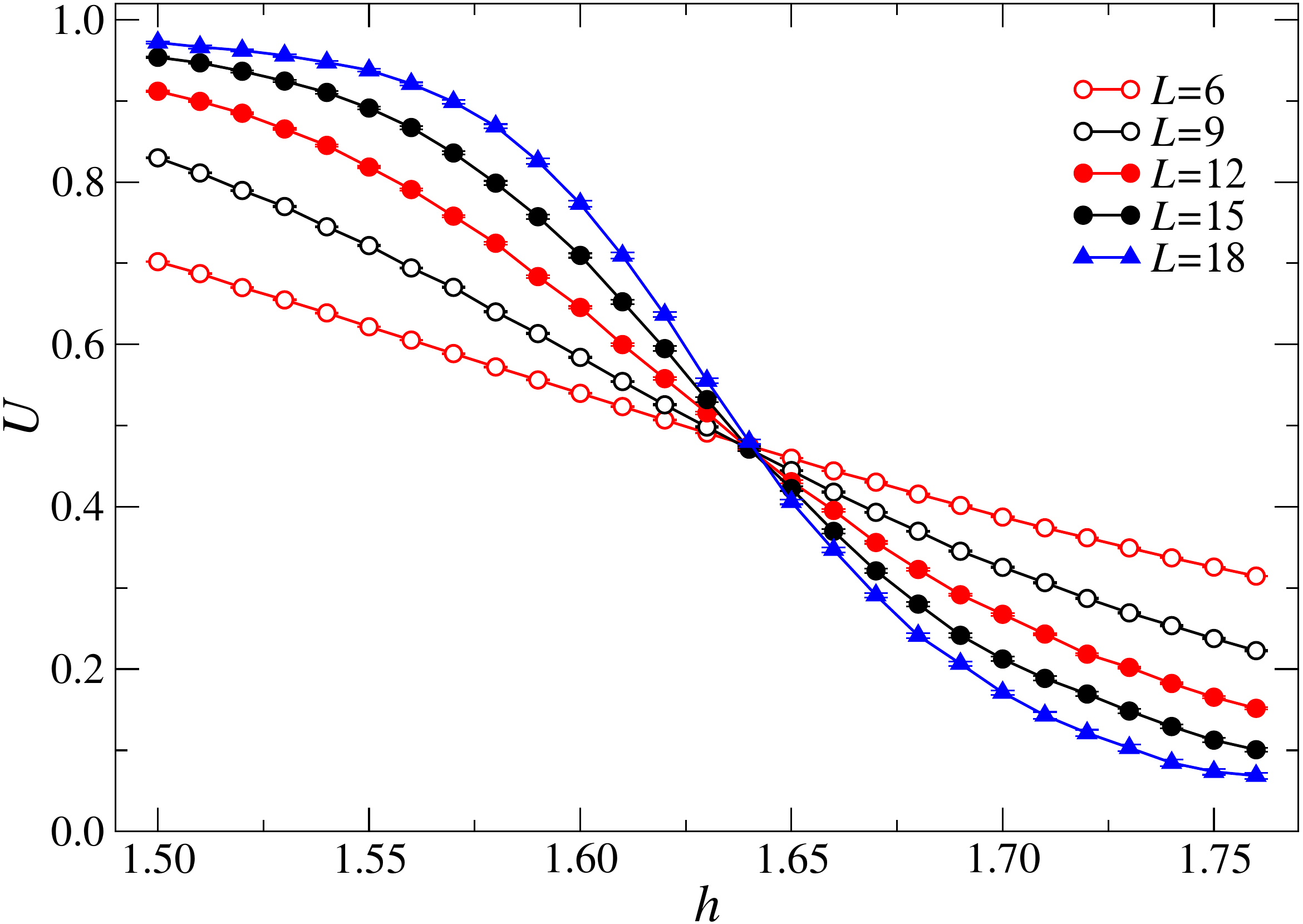}
\centering
\caption{(color online) Binder cumulant for different system sizes for the triangle lattice TFFIM. As $L$ increases, $U$ is monotonically becoming a step function -- indicating a continous phase transition at $h_c=1.64(1)$.}
\label{fig:bindr-t}
\end{figure}

The precise position of the QCP can be determined from the crossing point of the Binder cumulant in Fig.~\ref{fig:bindr-t}. With system size up to $L=18$, we can determine $h_{c}=1.64(1)$, with much higher accuracy than the previous determined values~\cite{Isakov2003}. From Fig.~\ref{fig:bindr-t}, it is also interesting to notice that the Binder cumulant for the triangle lattice TFFIM behaves in a regular manner, as system size $L$ increases: the Binder cumulant turns to be closer towards a step function -- meaning the QPT at $h_c$ is indeed a continuous phase transition, i.e. a QCP.

\subsection{Honeycomb lattice}
\label{sec:results_honeycomb}
\begin{figure}[htp!]
\centering
\includegraphics[width=\columnwidth]{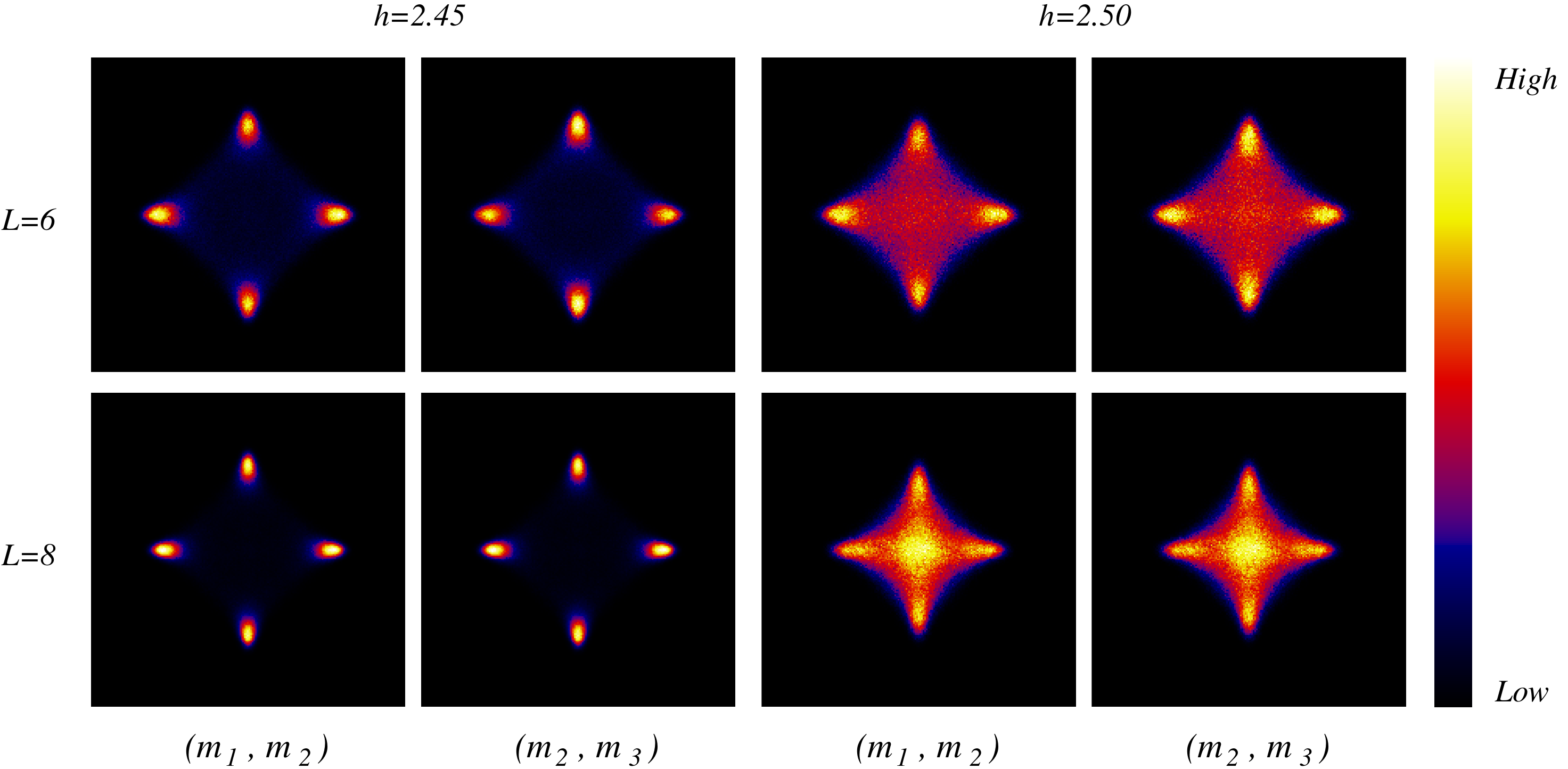}
\caption{(color online) Upper row: histograms of the order parameter for the honeycomb lattice TFFIM at $h=2.45$ and $h=2.5$ for $L=6$. Lower row: histogram of the order parameter for the honeycomb lattice TFFIM with same set of $h$ for $L=8$. There is {\it NO} sign of emergent continuous $O(3)$ symmetry. Note, according to the crossing of the Binder cumulant in Fig.~\ref{fig:bindr-h}, the quantum phase transition is at $h_c=2.48(1)$. The histograms at $h=2.5$ clearly contain the coexistence of the discrete symmetry breaking at $h<h_c$ and the zero-magnetization at $h>h_c$.}
\label{fig:histogram_honeycomb}
\end{figure}

\begin{figure}[htp!]
\includegraphics[width=\columnwidth]{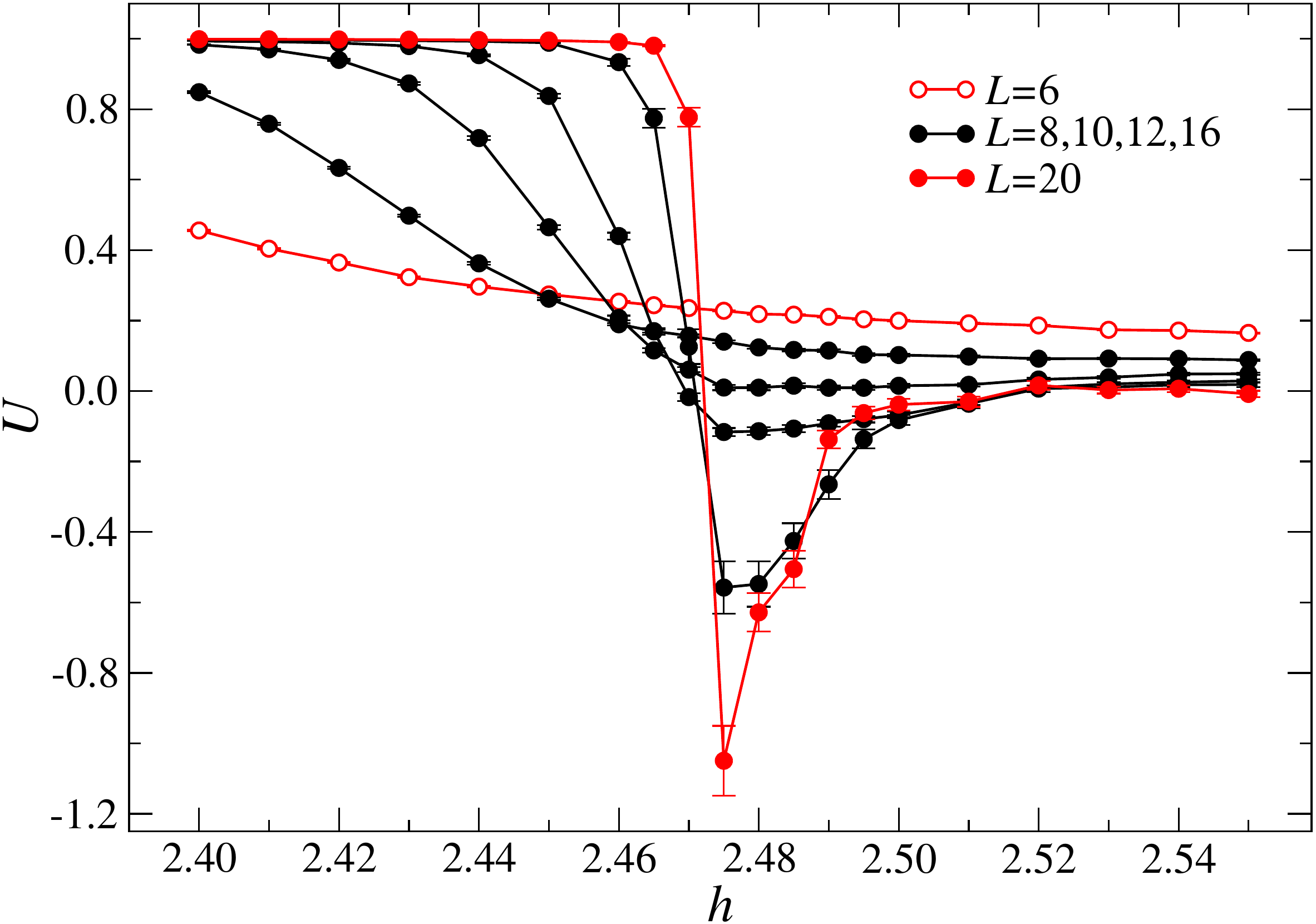}
\centering
\caption{(color online) Binder cumulant for the honeycomb lattice TFFIM. As system size increases, it is clear that $U$ develops a negative peak that grows with increasing $L$ -- indicating a first order phase transition at $h=2.48(1)$.}
\label{fig:bindr-h}
\end{figure}

Fig.~\ref{fig:histogram_honeycomb} shows the order parameter histogram of the honeycomb lattice TFFIM. Since the order parameter is a $3$D vector, we depict the histogram in two $2$D cuts: $(m_1,m_2,m_3=0)$ and $(m_1=0,m_2,m_3)$. The upper row is the data for smaller system size $L=6$ and the lower row is the data for larger system size $L=8$. At transverse field $h=2.45<h_c$, the histogram show discrete points, at the position of $m_1=\pm\frac{m}{\sqrt{3}}$, $m_2=\pm\frac{m}{\sqrt{3}}$ and $m_3=\pm\frac{m}{\sqrt{3}}$, which means the system is inside the discrete symmetry breaking phase with 6-fold degeneracy. Actually $h=2.45$ is already very close to the QPT at $h_c=2.48(1)$ (determined by the Binder cumulant in Fig.~\ref{fig:bindr-h}). For finite size system, even when we go slightly above the thermodynamic $h_c$, as shown in the right panel in Fig.~\ref{fig:histogram_honeycomb} with $h=2.5$, the discrete points in the histogram are still clearly presented, and the histogram counts in the center of the chart ($m_1=m_2=m_3=0$) also starts to increase. This means that the QPT at the honeycomb lattice TFFIM is different from the triangle lattice case, in that, it {\it does not} develop an emergent continuous $O(3)$ symmetry, as in the theoretical proposal of Ref.~\cite{Roychowdhury2015}. Instead, the discrete symmetry breaking persists all the way to the QPT point. The coexistence of maximums in the histograms at both the discrete points and the point in the center is a hallmark of a first-order phase transition, it also hints (will be explained in Sec.~\ref{sec:anisotropy}) that in the QPT of the honeycomb lattice TFFIM, the anisotropic terms in the effective Hamiltonian (Eq.~\ref{eq:h_lgw2}) play an important role in understanding the nature of the phase transition.

Again, the precise position of the QPT in the honeycomb lattice TFFIM is determined by the Binder cumulant defined as Eq.~\ref{eq:binder-h}. The results are shown in Fig.~\ref{fig:bindr-h}. Different systems also cross at a single point, and the position is the $h_c=2.48(1)$. However, as the system sizes increases, one observes that instead of becoming a step function at $h_c$, the Binder cumulant becomes narrower and has a tendency towards negatively diverging values from $L=12$ to $L=20$. This signals that it is clearly not a continuous phase transition and it is consistent with a first-order phase transition~\cite{Jin2012}.

\onecolumngrid

\twocolumngrid

From the comparison of MC results in order parameter histogram and Binder cumulant, it is now obvious that the emergent $O(2)$ symmetry is present in the triangle lattice TFFIM, but the anticipated emergent $O(3)$ symmetry is absent in the honeycomb lattice TFFIM. In the next Section (Sec.~\ref{sec:anisotropy}), we unveil the reason behind such a contrast.

\subsection{Measuring anisotropy in the effective model}
\label{sec:anisotropy}

In this section we derive the method of directly measuring the anisotropic terms in the effective models in Eqs.~\eqref{eq:effectiveHamiltonianT} and \eqref{eq:h_lgw2}, and present the corresponding MC data to elucidate the reason behind the presence/absence of emergent continuous symmetry in the triangle/honeycomb TFFIMs.

\begin{figure}[htp!]
\includegraphics[width=\columnwidth]{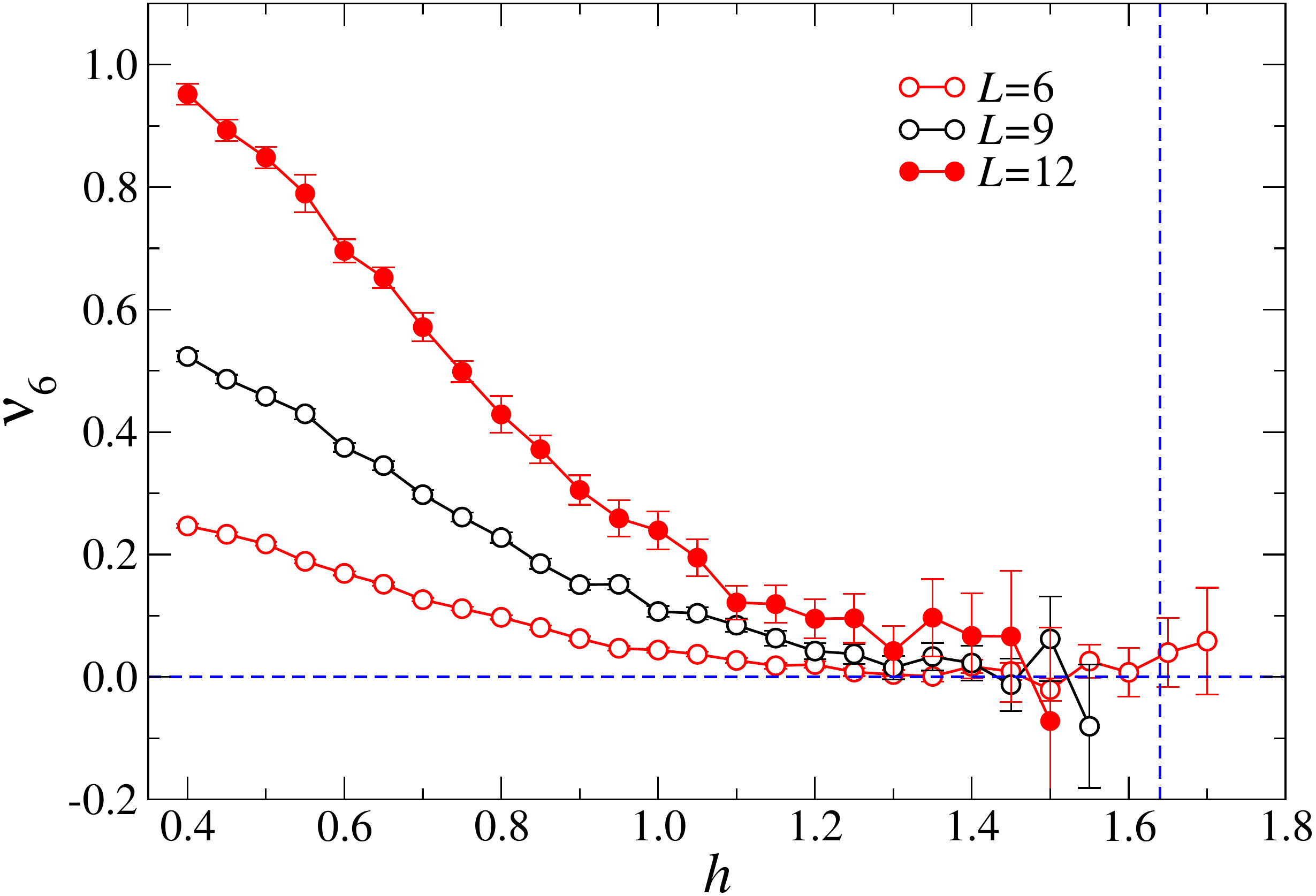}
\centering
\caption{(color online) Anisotropic term $\nu_6$ as function of transverse field $h$
 for different system sizes on the triangle lattice TFFIM. $\nu_6$ goes to zero at the QCP, giving rise to the emergent continuous $O(2)$ symmetry. The blue vertical dash line highlights the position of $h_c$. The blue horizontal dash line highlights the value of $\nu_6=0$.}
\label{fig:c6-t}
\end{figure}

\begin{figure}[h!]
\includegraphics[width=\columnwidth]{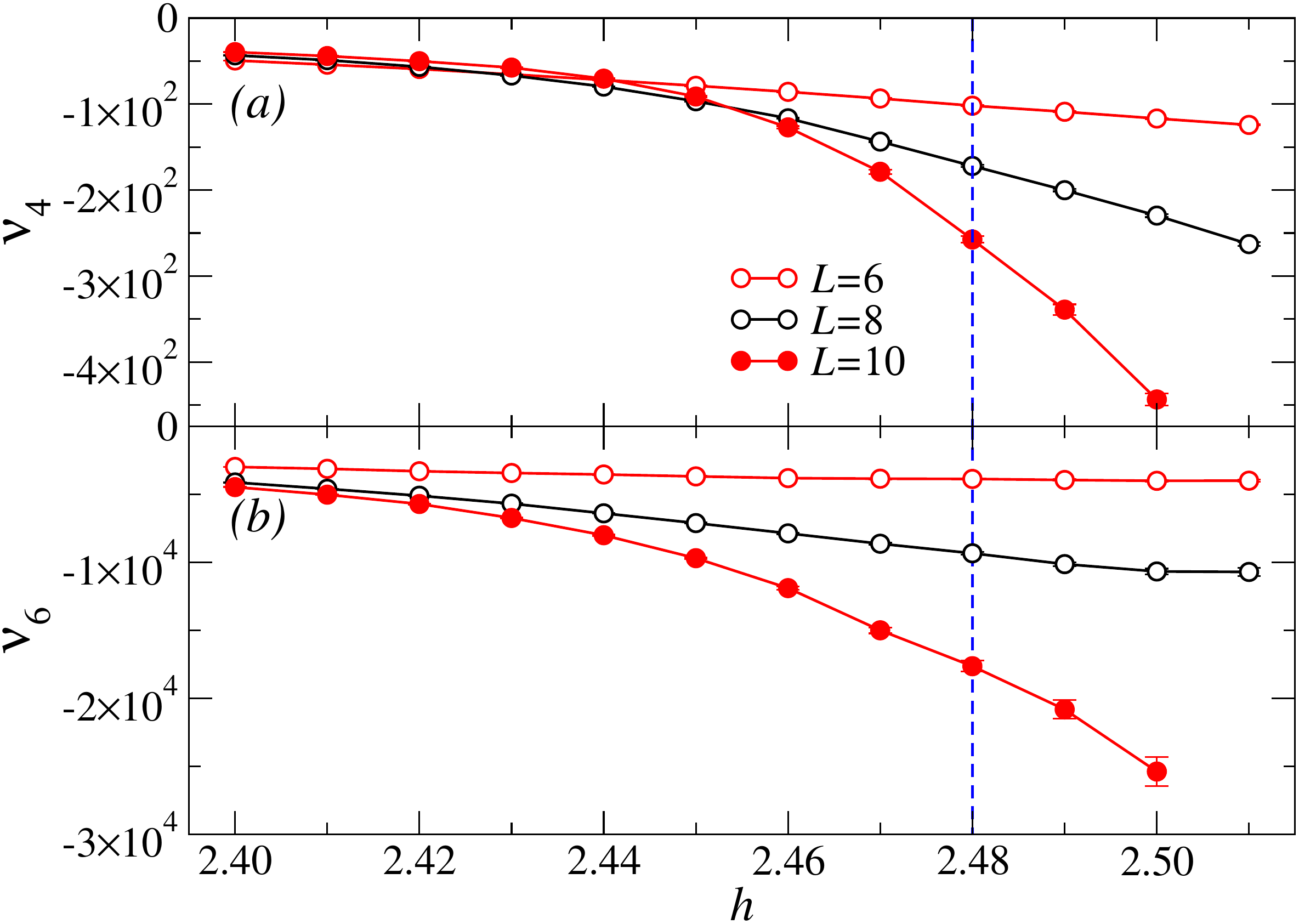}
\centering
\caption{(color online) $\nu_4$ (a) and $\nu_6$ (b) as function of transverse field $h$
 for different system size on the honeycomb lattice TFFIM. As $h$ is approaching the $h_c$, $\nu_4$ and $\nu_6$ increase greatly, in obvious contrast to the anisotropic term in the triangle lattice case (see Fig.~\ref{fig:c6-t}), and give rise to a first order phase transition (see Fig.~\ref{fig:histogram_honeycomb} and Fig.~\ref{fig:bindr-h}). The blue vertical dash line highlights the position of $h_c$.}
\label{fig:a4b6}
\end{figure}

We begin with the effective model in Eq.~\eqref{eq:effectiveHamiltonianT}, where the anisotropy, to the leading order, is represented by the term proportional to $\nu_6$. To extract the coefficient $\nu_6$ from Monte Carlo simulations, we consider the following expectation value,
\[\left\langle\cos(6\theta)\right\rangle
    =\frac1Z\sum_{\{\sigma_i\}}\cos(6\theta)e^{-H_{LGW}^T}.\]
To evaluate this average, we separate $H_{LGW}^T$ into two terms: $H_{LGW}^T=H_0+\nu_6m^6\cos(6\theta)$, where $H_0$ does not depend on $\theta$. Then, to the leading order of $\nu_6$, we can expand the exponential function and get
\begin{align*}
  \left\langle\cos(6\theta)\right\rangle
  &=\frac1Z\sum_{\{\sigma_i\}}\cos(6\theta)e^{-H_0}
  \left[1-\nu_6 m^6\cos(6\theta)+\cdots\right]\\
  &=\left<\cos(6\theta)\right>_0
    -\nu_6\left<m^6\cos^2(6\theta)\right>_0+\cdots,
\end{align*}
where $\langle\rangle_0$ denotes averages under Boltzmann weights determined by $H_0$. Since $H_0$ is isotropic in $\theta$, the average $\langle\cos(6\theta)\rangle_0$ vanishes and $\langle m^6\cos^2(6\theta)\rangle_0=\langle m^6\rangle/2$. Therefore, the equation above can be simplified to
\begin{equation}
  \label{eq:avg6theta}
  \langle\cos(6\theta)\rangle=-\frac{\nu_6}2\left<m^6\right>+\cdots.
\end{equation}
Hence, the anisotropy coefficient $\nu_6$ can be determined from Monte Carlo simulations using
\begin{equation}
  \label{eq:nu6mc}
  \nu_6=-\frac{2\langle\cos(6\theta)\rangle}{\left<m^6\right>}.
\end{equation}
From the order parameter histogram, for each configuration, we can read of the $\theta$ angle according to Eq.~\ref{eq:op-t}, and readily obtain the $\langle\cos(6\theta)\rangle$ via the MC configuration average. The 6th moment of $\langle m^6 \rangle$ can also be measured from the simulations. Hence the $\nu_6$ is obtained, and as shown in Fig.~\ref{fig:c6-t}.

Clearly, as $h$ approaches $h_c$, the anisotropic term systematically goes to zero, for all different system sizes studied. In the contrary, it is finite deep in the ordered phase for $h<1.2$, and the coefficient $\nu_6$ grows with the system size. In Refs.~\cite{Moessner2001,Isakov2003}, it is argued that the anisotropy term is irrelevant near the $O(2)$ fixed point, and as a result, it is irrelevant at the critical point, which then has an emergent $O(2)$ symmetry. Our numerical simulation demonstrates both its absence at the critical point and its relevance deep in the ordered phase.

The situation is very different for the TFFIM on the honeycomb lattice. First of all, the anisotropy parameters $\nu_4$ and $\nu_6$ in the effective model in Eq.~\eqref{eq:h_lgw2} can also be determined from the Monte Carlo simulations. As discussed in Sec.~\ref{sec:orderparameterhisto}, one can parameterize $m_{1,2,3}$ using spherical coordinates $m$, $\theta$ and $\phi$: $m_1=m\cos\theta$, $m_2=m\sin\theta\cos\phi$ and $m_3=m\sin\theta\sin\phi$.
Next, we consider the averages of two spherical harmonics $Y_4^0$ and $Y_6^0$:
\begin{align}
  \label{eq:y40}
  Y_4^0&=\frac3{16\sqrt\pi}\left(3-30\cos^2\theta+35\cos^4\theta\right)\\
  \label{eq:y60}
  Y_6^0&=\frac{\sqrt{13}}{32\sqrt\pi}\left(-5+105\cos^2\theta-315\cos^4\theta
  +231\cos^6\theta\right).
\end{align}
Expanding the effective Hamiltonian in Eq.~\eqref{eq:h_lgw2} to the leading order of $\nu_4$ and $\nu_6$, we get
\begin{widetext}
  \begin{align*}
    \left<Y_4^0\right>&=\left<Y_4^0\right>_0
                        -\nu_4\left<Y_4^0(m_1^4+m_2^4+m_3^4)\right>_0
                        -\nu_6\left<Y_4^0(m_1m_2m_3)^2\right>_0
                        =-\frac{1}{15\sqrt{\pi}}\nu_4\left<m^4\right>_0
                        -\frac{1}{330\sqrt\pi}\nu_6\left<m^6\right>_0,\\
   \left<Y_6^0\right>&=\left<Y_6^0\right>_0
    -\nu_4\left<Y_6^0(m_1^4+m_2^4+m_3^4)\right>_0
    -\nu_6\left<Y_6^0(m_1m_2m_3)^2\right>_0
    =-\frac{1}{231\sqrt{13\pi}}\nu_6\left<m^6\right>_0.
  \end{align*}
\end{widetext}
Using these results, we can determine $\nu_4$ and $\nu_6$ from Monte Carlo simulations as,
\begin{align}
  \nu_4&=-\frac{15\sqrt\pi}{\langle m^4\rangle}
         \left(\langle Y_4^0\rangle-\frac{7\sqrt{13}}{10}\langle Y_6^0\rangle\right)
     \label{eq:nu4mc2} \\
  \nu_6&=-\frac{231\sqrt{13\pi}}{\langle m^6\rangle}\langle Y_6^0\rangle.
\label{eq:nu6mc2}
\end{align}

For each configuration in the order parameter histogram, we can determine the $\theta$ angle and hence obtain the expectation values $\langle Y_4^0 \rangle$ and $\langle Y_6^0 \rangle$. Then the arrive at $\nu_4$ and $\nu_6$ from Eq.~\ref{eq:nu4mc2} and ~\ref{eq:nu6mc2}.

Our simulation results in Fig.~\ref{fig:a4b6} show that near the critical point (for $2.44<h<2.50$), both $\nu_4$ and $\nu_6$ are finite, and their values grow with the system size. These results imply that they are both relevant perturbations in the effective LGW theory. The presence of these anisotropic terms explains the lack of an emergent $O(3)$ symmetry at the phase transition. Furthermore, we notice that the value of $\nu_4$ extracted is negative, and this is related to the fact that the QPT is first-order. Along one particular radial direction in the parameter space $(m_1, m_2, m_3)$, the angles $\theta$ and $\phi$ are fixed, and the LGW effective potential is a function of $m$,
\begin{equation}
  \label{eq:lgw-dir}
    H_{LGW}=\left[u_4+\nu_4f_4(\theta, \phi)\right]m^4
    +\left[u_6+\nu_6f_6(\theta, \phi)\right]m^6
    +\cdots,
\end{equation}
where the angular dependent functions $f_4(\theta,\phi)=\cos^4\theta+\sin^4\theta\cos^4\phi +\sin^4\theta\sin^4\phi$ and $f_6(\theta, \phi)=\cos^2\theta\sin^4\theta\cos^2\phi\sin^2\phi$. It is well-known that in a LGW effective potential, a negative quartic term results in a first-order phase transition, at which the effective potential has two minimums, one of which at $m=0$. From Fig.~\ref{fig:histogram_honeycomb}, one can determine that at $h=2.50$, the effective potential has two minimums along the directions of $m_1=m_2=0$, $m_2=m_3=0$, and $m_1=m_3=0$, but only one minimum along the diagonal directions. Hence, the quartic term is negative along the directions of $m_i=m_j=0$, and positive along the diagonal directions. This is consistent with our finding of $\nu_4$ being negative, which implies that the coefficient of the quartic term, $u_4+\nu_4f_4(\theta,\phi)$, is smaller (more negative) along the directions of $m_i=m_j=0$, where $f_4(\theta,\phi)$ is maximal. In summary, a large and negative anisotropic term $\nu_4$ explains that the histograms in Fig.~\ref{fig:histogram_honeycomb} are anisotropic, and have more than one maximum at the phase transition, which in turn indicates that the QPT is first-order.

The Refs.~\cite{Roychowdhury2015} and \cite{XuVBS2011} had contradicting conclusions on whether a QPT described by the LGW effective potential in Eq.~\eqref{eq:h_lgw2} can be second-order, and the key issue behind that is whether the anisotropic term $\nu_4$, known as the cubic anisotropy since it respects the cubic symmetry in the parameter space, is relevant or irrelevant. Although early studies based on leading-order $\epsilon$-expansion calculations suggests its irrelevance~\cite{Aharony1973, Ketley1973, Brezin1974}, a later six-loop calculation~\cite{Calabrese2003} shows that it is indeed relevant. However, we notice that the obtained scaling dimension is very close to zero, and the estimated error is of the same order as the magnitude of the scaling dimension.

Our finding is consistent with Ref.~\cite{XuVBS2011}, which argues that a negative cubic anisotropy is relevant and will make the QPT first-order, and it is contrary to the assumption in Ref.~\cite{Roychowdhury2015}, which suggests that the honeycomb lattice TFFIM realizes a continuous QCP where both $\nu_4$ and $\nu_6$ terms are irrelevant perturbations, and as a result, the QCP as an emergent O(3) symmetry. However, we note that our numerical simulation does not rule out the possibility that the scheme in Ref.~\cite{Roychowdhury2015} is still correct and just not realized in this particular model. The scaling dimension of the cubic anisotropic term at the (2+1)D O(3) Wilson-Fisher fixed point can be studied by further numerical studies of the correlation function of such anisotropic terms at a QCP with the $O(3)$ universal class.

We further notice that the derivations in this section rely on expansions with respect to the anisotropic coefficients $\nu_4$ and $\nu_6$, and the numbers obtained from Monte Carlo simulations are only quantitatively correct when the anisotropies are small. However, the observed behaviors of large anisotropies in certain parameter ranges are still qualitatively correct.

\section{SUMMARY AND OUTLOOK}
\label{sec:summary}

In this work, we study the QPTs in the triangle lattice and the honeycomb lattice TFFIM, using large-scale Monte Carlo simulations. In particular, we evaluate the anisotropic terms related to the proposed emergent continuous symmetry, in the low-energy effective models. Our simulation confirms that the QPT in the triangle lattice TFFIM is second-order, and the anisotropic term is irrelevant at the QPT, resulting an emergent $O(2)$ symmetry. However, our simulation reveals that the QPT in the honeycomb lattice TFFIM is first-order. Furthermore, the calculated anisotropic terms remain finite and hence behave as relevant perturbations at the QPT. This indicates that it is the irrelevant/relevant of the anisotropic terms in the effective Hamiltonian that gives rise an emergent continuous symmetry QCP in the triangle TFFIM but a first order QPT in the honeycomb TFFIM.

Our numerical study in this work set an example of careful and controlled investigation of the low-energy effective quantum field theory in frustrated magnetic systems. The method of order-parameter histogram developed in this work is a generic approach, and it can be applied to other models to compare numerical simulations to theoretical analysis of emergent symmetries and the relevance of perturbations based on LGW-type analysis of low-energy effective theories. Since such type of theoretical analyses are widely used in these days in analyzing and proposing novel properties of classical and quantum phase transitions, a more scrutinizingly approach, such as the one employed in this work, can give more solid evidences.

\begin{acknowledgments}

We are in debt to S. Bhattacharjee and F. Pollmann for bringing our attention to this problem and for the stimulating discussions over the project. We thank L. Balents and C. Xu for pointing out the references~\cite{Calabrese2003,XuVBS2011}. We acknowledge C.-X. Ding and Y.-J. Deng for sharing the knowledge of the order parameter histogram and the geometric cluster update scheme. We also acknowledge W.-A. Guo for discussing the usage of Binder cumulant to detect first order phase transition. The numerical calculations were carried out at the supercomputing platforms in the Center for Quantum Simulation Sciences in the Institute of Physics, Chinese Academy of Sciences as well as the National Supercomputer Center in Tianjin on the Tianhe-1A platform. YCW and ZYM are supported by the Ministry of Science and Technology of China through National Key Research and Development Program under Grant No. 2016YFA0300502, National Natural Science Foundation of China (NSFC Grant Nos. 11421092, 11574359 and 11674370) and the National Thousand-Young-Talents Program of China. YQ is supported in part by Perimeter Institute for Theoretical Physics. Research at Perimeter Institute is supported by the Government of Canada through Industry Canada and by the Province of Ontario through the Ministry of Research and Innovation. SC acknowledges the support from NSFC under Grants Nos. 11425419, 11374354 and 11174360.
\end{acknowledgments}

\bibliography{IsingHoneycomb}

\end{document}